\documentclass[aps,prx,reprint,superscriptaddress]{revtex4-2}
 
\usepackage{graphicx}
\usepackage{subfigure}
\usepackage{amsmath}
\usepackage{amsfonts}
\usepackage{amssymb}
\usepackage{times,txfonts}
\usepackage{braket}
\usepackage[colorlinks=true,linkcolor=blue,urlcolor=blue,citecolor=blue]{hyperref}

\begin{document}

\title{Universal energy fluctuations in inelastic scattering processes}

\author{Samuel L. Jacob}
\email{samjac91@gmail.com}
\affiliation{School of Physics, Trinity College Dublin, Dublin 2, Ireland}

\author{John Goold}
\email{gooldj@tcd.ie}
\affiliation{School of Physics, Trinity College Dublin, Dublin 2, Ireland}
\affiliation{Trinity Quantum Alliance, Unit 16, Trinity Technology and Enterprise Centre, Pearse Street, Dublin 2, D02 YN67, Ireland}
\affiliation{Algorithmiq Limited, Kanavakatu 3C 00160 Helsinki, Finland}

\date{\today}
\author{Gabriel T. Landi}
\affiliation{Department of Physics and Astronomy, University of Rochester, Rochester, New York 14627, USA}
\email{glandi@ur.rochester.edu}

\author{Felipe Barra}
\email{fbarra@dfi.uchile.cl }
\affiliation{Departamento de F\'isica, Facultad de Ciencias F\'isicas y Matem\'aticas, Universidad de Chile, 837.0415 Santiago, Chile}

\begin{abstract}

Quantum scattering  is used ubiquitously in both experimental and theoretical physics across a wide range of disciplines, from high-energy physics to mesoscopic physics. In this work, we uncover universal relations for the energy fluctuations of a quantum system scattering inelastically with a particle at arbitrary kinetic energies. In particular, we prove a fluctuation relation describing an asymmetry between energy absorbing and releasing processes which relies on the non-unital nature of the underlying quantum map. This allows us to derive a bound on the average energy exchanged. We find that energy releasing processes are dominant when the kinetic energy of the particle is comparable to the system energies, but are forbidden at very high kinetic energies where well known fluctuation relations are recovered. Our work provides a unified view of energy fluctuations when the source driving the system is not macroscopic but rather an auxiliary quantum particle in a scattering process. 
\end{abstract}

\maketitle{}

\textit{Introduction. ---} Scattering is a mechanism of interaction between physical systems that is pervasive across nature and experiment, from low to high energies \cite{Belkic2004,Taylor2006}. It is an essential tool in the characterization of materials and quantum phenomena in condensed matter \cite{Furrer2009,Bramwell2014}, in describing the transport properties of quantum systems \cite{Buttiker1992,Emberly2000,Imry2005,Nazarov2009,Moskalets2012,Benenti2017} and the properties of ultracold gases \cite{Inguscio1999,Hohmann2016,Schmidt2018,Schmidt2019,Widera2020,Irfan2021,Bouton2021,Adam2022,Nettersheim2023}. Quantum scattering theory describes how two (or more) quantum systems change their state after they collide, which entails an energy exchange between them when the scattering process is inelastic \cite{Belkic2004,Taylor2006,Furrer2009}. Such energy exchanges have been recently analysed from a thermodynamic viewpoint, showing that a particle colliding with a quantum system can act as a source of heat \cite{Jacob2021,Tabanera2022,Tabanera2023} or work \cite{Jacob2022}. Very massive particles can also be used to probe the energy statistics that would result from a two-point measurement scheme on the system \cite{Jacob2023}. Although these studies validate scattering as a powerful microscopic approach to thermodynamics of quantum systems, a more general treatment at the level of energy fluctuations is still not available.


In thermodynamics, energy fluctuations are usually studied for small -- classical or quantum -- systems interacting with macroscopic sources. The assumption of a macroscopic source allows us to define some Hamiltonian for the system with time dependent parameter that we imagine is operated in a classical way \cite{Balian1991}. Within this paradigm, some of the most famous results of stochastic thermodynamics have been derived, for example the so-called fluctuation relations \cite{Evans1993,Gallavotti1995,Crooks1999,Esposito2009,Campisi2011,Jarzynski2011}. As an example, consider a system of any size prepared in thermal equilibrium with its environment characterized by $\beta = 1/k_B T$, where $k_B$ is the Boltzmann constant and $T$ is the temperature. When the system is driven out of equilibrium by a macroscopic source in a cyclic fashion (so that the system Hamiltonian is the same before and after the interaction), then the fluctuation relation reads $e^{-\beta W}p_W = \tilde{p}_{-W}$, where $p_{W}$ is the probability distribution for an energy change $W$ during the process and $\tilde{p}_{-W}$ is the probability distribution for an energy change $-W$ in the time-reversed process \cite{Esposito2009,Campisi2011}. Jarzynski's equality $\langle e^{-\beta W} \rangle = 1$ \cite{Jarzynski1997} follows by a simple average over $W$ which, through Jensen's inequality, implies (on average) the impossibility of energy extraction in a cyclic process $\langle W \rangle \geq 0$. Since the macroscopic source is considered to behave deterministically, i.e. as a work source, the energy consumed is interpreted as work done on the system. Through the use of the two-point measurement scheme, fluctuation relations have been extended to closed quantum systems \cite{Kurchan2000,Tasaki2000,Talkner2007,Talkner2007a,Hanggi2015,Binder2018}, derived for macroscopic heat sources \cite{Jarzynski2004,Esposito2009} and experimentally verified across different platforms \cite{An2014,Smith2018,Micadei2021,Solfanelli2021,Aguilar2022}. 

A valid framework to go beyond the macroscopic source paradigm is that of open quantum systems \cite{Breuer2007,Rivas2012}, where the system dynamics is described by a dynamical map \cite{Kraus1971,Kraus1983,Nielsen2000,Alicki2007} obtained after the interaction with another quantum system of arbitrary size. It is known that fluctuation relations can be derived within the two-point measurement scheme when the map is unital, i.e. if the maximally-mixed state is an invariant state \cite{Rastegin2013,Rastegin2014,Smith2018}. If the map is not unital, it has been shown that Jarzynski's equality is modified to $\langle e^{-\beta W} \rangle = 1 + \eta$, where $\eta$ can be positive or negative \cite{Rastegin2014,Goold2015,Rastegin2018,Goold2021}. This generalized equality has been experimentally verified with entangled photons subject to turbulence \cite{Ribeiro2020} and also appears in studies of fluctuations with generalized measurements \cite{Watanabe2014} and feedback control \cite{Sagawa2010,Morikuni2011}. Since Jensen's inequality then implies $\langle W \rangle \geq -\beta^{-1}\log(1 + \eta)$, this suggests that $\eta > 0$ allows for energy releasing processes; indeed, such processes are necessary for cooling quantum systems \cite{Binder2018}. Despite their clear relevance for thermodynamics, the physics behind non-unital fluctuations remains poorly understood and appreciated. Arguably, this is due to the fact that previous studies \cite{Rastegin2014,Goold2015,Rastegin2018,Goold2021} focus on the properties of the dynamical map rather than on a quantum mechanical description of the interacting systems. Progress could be made by using a realistic and microscopic approach like quantum scattering theory, whereby one treats the interacting systems as quantum systems in their own right, potentially providing a unified view of energy fluctuations beyond the macroscopic source limit.



In this Letter, we provide such a unified view on energy fluctuations by studying a quantum system scattering inelastically with a particle at an arbitrary kinetic energy. The dynamical map for the system naturally encodes its energy fluctuations without relying on any measurement scheme. Our main result [Eq.~\eqref{fluctuationrelation}] describes a universal fluctuation relation obeyed for a system driven out of equilibrium by the colliding particle and reflects the non-unital nature of the scattering process. From this result, we derive an exact bound for the energy exchanged [Eqs.~\eqref{inequality} and \eqref{eta}] as a function of the particle's kinetic energy. We show that non-unitality dominates when the particle's kinetic energy is comparable to the energy gaps of the system, allowing energy extraction from the system; while at very high kinetic energies we recover unitality and the standard fluctuation relation. Our results show that non-unital fluctuation relations are intimately connected with the energy of the quantum source, which can be of the same order as energy fluctuations themselves.

\textit{Setup and energy fluctuations. ---} We consider a quantum scattering process between a system $S$ and a particle $P$. In a reference frame
co-moving with the center of mass, only the reduced
mass plays a role, but we simplify the treatment
by fixing the position of system $S$ and consider the particle $P$ to be travelling in one direction with associated momentum $\hat{p}$ and position $\hat{x}$ operators. The total Hamiltonian is $\hat{H} = \hat{H}_0 + \hat{V}(\hat{x})$ where $\hat{H}_0 = \hat{H}_S \otimes \hat{\mathbb{I}}_P + \hat{\mathbb{I}}_S \otimes \hat{p}^2/2m$ is the bare Hamiltonian. The energy of the system is defined by $\hat{H}_S \ket{j}=e_j\ket{j}$, where $\{ \ket{j} \}$ is a basis of eigenvectors associated to its discrete energy spectrum $\{ e_j \}$. The energy of the particle is described by $\hat{p}^2/2m \ket{p} = E_p \ket{p}$, where $\{ \ket{p} \}$ are improper (non-normalizable) eigenvectors whose position representation are plane waves $\braket{x|p} = \exp( i p x / \hbar) / \sqrt{2 \pi \hbar}$ and $E_p = p^2 / 2m \geq 0$ is the kinetic energy. The interaction operator $\hat{V}(\hat{x})$ is assumed to vanish sufficiently far away from the scattering region where the system is located, so that the unitary scattering operator $\hat{S} = \lim_{t\to+\infty} e^{\frac{it}{\hbar}\hat{H}_0}e^{-\frac{i2t}{\hbar}\hat{H}}e^{\frac{i t}{\hbar}\hat{H}_0}$ exists and satisfies energy conservation $[\hat{S},\hat{H}_0]=0$ \cite{Belkic2004,Taylor2006}. Considering the initial state of the system $\hat{\rho}_S$ and particle $\hat{\rho}_P$ to be uncorrelated before the collision, the state of the system after the collision is
\begin{align}
    \label{map}
    \Phi(\hat{\rho}_S) = \mathrm{Tr}_P[\hat{S}(\hat{\rho}_S \otimes \hat{\rho}_P) \hat{S}^{\dagger}] \; ,
\end{align}
where $\mathrm{Tr}_P$ is the partial trace over the particle and $\Phi$ is a completely positive and trace preserving map \cite{Breuer2007,Rivas2012,Kraus1971,Kraus1983,Alicki2007}. 

\begin{figure}[t!]
\centering
\includegraphics[width=0.48\textwidth]{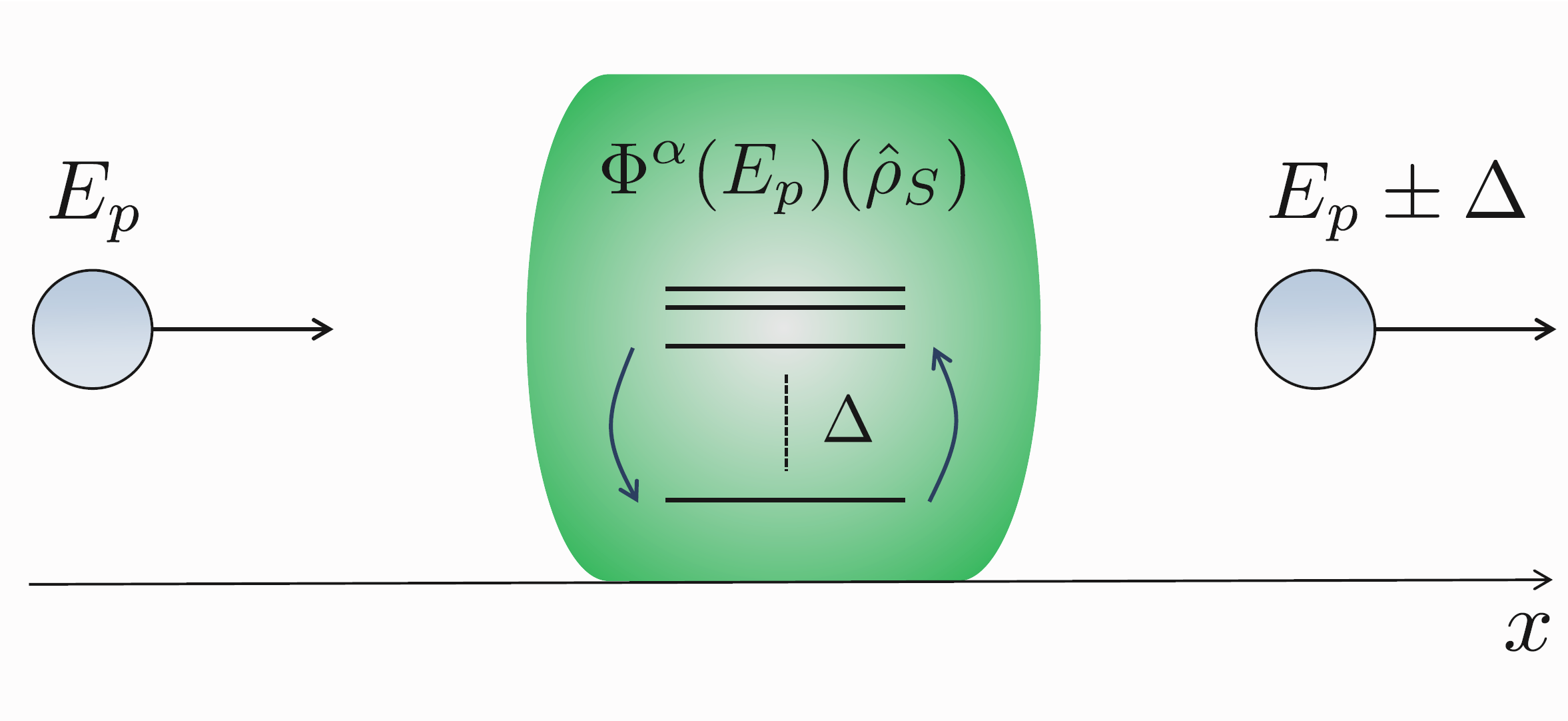}
\caption{\label{fig-scheme} A particle with kinetic energy $E_p$ travels in space with direction $\alpha$ and scatters with the system initially in state $\hat{\rho}_S$. The map $\Phi^{\alpha}(E_p)$ encodes the evolution and energy fluctuations of the system as defined in Eqs.~\eqref{scatteringmap} and \eqref{scatteringmapoperation}. The direction of the particle is to the left $\alpha = +$ in this schematic for illustrative purposes.}
\end{figure}

The explicit evaluation of Eq.~\eqref{map} can be performed in the following kinetic energy eigenstates $\ket{E^{\alpha}_{p}} \equiv \sqrt{m/|p|} \ket{p}$, where $\alpha = \mathrm{sign}(p)$ accounts for the initial direction of the incoming particle, which can be travelling to the left ($\alpha = +$) or right ($\alpha = -$). First, we need the representation of the scattering operator in this basis which reads $\braket{E^{\alpha'}_{p'}|\hat{S}|E^{\alpha}_{p}} = \sum_{j',j} \ket{j'}\bra{j} \braket{E^{\alpha'}_{p'},j'|\hat{S}|E^{\alpha}_{p},j}$ where $ \ket{E^{\alpha}_p,j}$ is the eigenbasis of $\hat{H}_0$ and $\braket{E^{\alpha'}_{p'},j'|\hat{S}|E^{\alpha}_{p},j} = \delta(E_{p'} + e_{j'} - E_p - e_j) s^{\alpha' \alpha}_{j'j}(E_p + e_j)$. In the last expression, the $\delta$ function ensures energy conservation for the collision and $s_{j'j}^{\alpha' \alpha}(E)$ is the scattering matrix encoding the transition amplitudes from $\ket{E^{\alpha}_p,j} \rightarrow \ket{E^{\alpha'}_{p'},j'}$ at total energy $E = E_p + e_j$ \cite{Belkic2004,Taylor2006}. Rewriting the sum over $j',j$ as a sum over energy differences $\Delta$ then yields simply $\braket{E^{\alpha'}_{p'}|\hat{S}|E^{\alpha}_{p}} = \sum_{\Delta} \delta(E_{p'} - E_{p} + \Delta) \hat{S}^{\alpha' \alpha}_{\Delta}(E_p)$ where
\begin{align}
    \label{eigenoperators}
    \hat{S}^{\alpha' \alpha}_{\Delta}(E_p) = \sum_{\substack{j',j:\\ e_{j'}-e_j = \Delta}} s_{j'j}^{\alpha' \alpha}(E_p + e_j) \ket{j'}\bra{j} \; ,
\end{align}
are eigenoperators of $\hat{H}_S$ and thus obey $[\hat{H}_S, \hat{S}^{\alpha' \alpha}_{\Delta}(E_p)] = \Delta \hat{S}^{\alpha' \alpha}_{\Delta}(E_p)$. Second, we need the representation of the particle's state in the same basis $\rho^{\alpha \beta}_P(E_{p},E_q) \equiv \braket{E^{\alpha}_{p}|\hat{\rho}_P|E^{\beta}_{q}}$ and we can carry out the trace in Eq.~\eqref{map}. After integrating the $\delta$ functions, we find that the particle's state becomes dependent on the energy differences as $\rho^{\alpha \beta}_P(E_{p},E_p - \Delta + \Delta')$ (see Appendix~\ref{app:derivationmap} for more details). However, as shown in Ref.~\cite{Jacob2021}, if the particle has a well-defined direction before the collision and is sufficiently narrow in kinetic energy with respect to the energy differences, then we can write
\begin{align}
    \label{narrow}
    \rho^{\alpha \beta}_P(E_{p},E_p - \Delta + \Delta') \simeq \delta_{\alpha \beta} \delta_{\Delta\Delta'}~ \rho^{\alpha}_P(E_{p}) \; ,
\end{align}
where $\rho^{\alpha}_P(E_p) \equiv \rho^{\alpha \alpha}_P(E_{p},E_p)$ is the kinetic energy distribution for a particle travelling with direction $\alpha$. In this case, Eq.~\eqref{map} can be written as
\begin{align}
    \label{scatteringmapfull}
    \Phi(\hat{\rho}_S) = \int dE_p~\sum_{\alpha=\pm}\rho^{\alpha}_P(E_p)~ \Phi^{\alpha}(E_p)(\hat{\rho}_S) \; ,
\end{align}
where $\Phi^{\alpha}(E_p)$ is a completely positive and trace preserving map conditioned on the particle's kinetic energy $E_p$ and direction $\alpha$ given by
\begin{align}
    \label{scatteringmap}
    \Phi^{\alpha}(E_p)(\cdot) & = \int dW~ \Phi^{\alpha}(E_p,W)(\cdot) \\
    \label{scatteringmapoperation}
    \Phi^{\alpha}(E_p,W)(\cdot) & = \sum_{\Delta}\delta(W-\Delta)\sum_{\alpha'}\hat{S}^{\alpha' \alpha}_{\Delta}(E_p)\cdot \hat{S}^{\alpha' \alpha}_{\Delta}(E_p)^{\dagger} \; .
\end{align}
Eqs.~\eqref{scatteringmapfull}, \eqref{scatteringmap} and \eqref{scatteringmapoperation} define the dynamical map (see Fig.~\ref{fig-scheme}). 

Note that the Kraus operators in Eq.~\eqref{scatteringmapoperation} are system eigenoperators due to condition \eqref{narrow}, inducing a transition with energy change $\Delta$. Indeed, assuming that $\hat{H}_S$ has a non-degenerate spectrum, it is easy to see that the quantum operation in Eq.~\eqref{scatteringmapoperation} defines a probability distribution for the energy changes through
\begin{align}
    \label{probabilitydistribution}
    P^{\alpha}(E_p,W) & = \mathrm{Tr}_S\big[\Phi^{\alpha}(E_p,W)(\hat{\rho}_S)\big] \nonumber \\
    & = \sum_{j',j} \delta(W-e_{j'}+e_{j}) P^{\alpha}_{j'j}(E_p + e_j) p_j \; ,    
\end{align}
where $p_j \equiv \braket{j|\hat{\rho}_S|j}$ and $P^{\alpha}_{j'j}(E_p + e_j) = \sum_{\alpha'} |s^{\alpha'\alpha}_{j'j}(E_p + e_j)|^2$ is the transition probability. Note that Eq.~\eqref{probabilitydistribution} has the same form as the distribution for energy changes induced by a unitary operator $U$ on the system in a two-point measurement scheme \cite{Tasaki2000,Kurchan2000,Talkner2007a,Esposito2009,Campisi2011}, with two crucial differences. First, there is no need for a two-point measurement scheme as a consequence of condition \eqref{narrow}: a particle with a well-defined kinetic energy effectively measures the energy changes in the system \cite{Jacob2023}. Second, the transition probabilities are dictated by $P^{\alpha}_{j'j}(E_p + e_j)$ instead of $|\braket{|j'|U|j}|^2$, thus becoming dependent on both the system and the particle's energy. The normalization $\int P^{\alpha}(E_p,W) dW = \sum_{j',j} P^{\alpha}_{j'j}(E_p + e_j) p_j = 1$ holds since the map in Eq.~\eqref{scatteringmap} is trace preserving by construction. Indeed, the property $\sum_{j'} P^{\alpha}_{j'j}(E_p + e_j) = 1$ can be proven independently from the unitarity of the scattering operator and holds for any fixed total energy $E$ (see Appendix~\ref{app:propertiesscatt}).

\textit{Main result. ---} We now take our system to be in a thermal state $\hat{\rho}_S = e^{-\beta \hat{H}_S}/Z$ where $Z = \mathrm{Tr}_S[e^{-\beta \hat{H}_S}]$ is the partition function \footnote{Strictly speaking, we do not require the system to be in a thermal state, but only that its populations are thermally distributed.}. Using the property of the eigenoperators $\hat{S}^{\alpha' \alpha}_{\Delta}(E_p) e^{-\beta \hat{H}_S} = e^{\beta \Delta} e^{-\beta \hat{H}_S} \hat{S}^{\alpha' \alpha}_{\Delta}(E_p)$, it is easy to see that the quantum operation satisfies $e^{-\beta W} \Phi^{\alpha}(E_p,W)(\hat{\rho}_S) = \hat{\rho}_S\Phi^{\alpha}(E_p,W)(\hat{\mathbb{I}}_S)$ and thus the distribution in Eq.~\eqref{probabilitydistribution} obeys
\begin{align}
    \label{fluctuationrelation}
    e^{-\beta W}P^{\alpha}(E_p,W) & = \mathbb{P}^{\alpha}(E_p,-W) \; ,    
\end{align}
where $\mathbb{P}^{\alpha}(E_p,-W)$ is dual probability distribution given by
\begin{align}
    \label{dualprobabilitydistribution}
    \mathbb{P}^{\alpha}(E_p,-W) & = \mathrm{Tr}_S\big[\Phi^{\alpha}(E_p,W)^{\dagger}(\hat{\rho}_S)\big] \nonumber \\
    & = \sum_{j',j} \delta(-W-e_{j}+e_{j'}) p_{j'} P^{\alpha}_{j'j}(E_p + e_j) \; ,    
\end{align}
with the dual operation defined by $\mathrm{Tr}_S\big[\hat{\rho}_S\Phi^{\alpha}(E_p,W)(\hat{\mathbb{I}}_S)\big] = \mathrm{Tr}_S\big[\Phi^{\alpha}(E_p,W)^{\dagger}(\hat{\rho}_S)\big]$. Eq.~\eqref{dualprobabilitydistribution} has the same form as the distribution for energy changes induced by a time-reversed unitary operator $U^{\dagger} = \Theta U \Theta^{\dagger}$ on the system in a two-point measurement scheme, where $\Theta$ is the (anti-unitary) time reversal operator \cite{Tasaki2000,Kurchan2000,Talkner2007a,Esposito2009,Campisi2011}. In this sense, the dual operation $\Phi^{\alpha}(E_p,W)^{\dagger}$ reverses the energy change induced by $\Phi^{\alpha}(E_p,W)$ \footnote{Note that the reverse energy change induced by the dual operation is generally not equivalent to time-reversing the collision as a whole. The latter requires a reversal of the particle direction and scattering matrix.}. However, a crucial point is that the dual distribution in Eq.~\eqref{dualprobabilitydistribution} is generally not normalized $\gamma^{\alpha}(E_p) \equiv \int \mathbb{P}^{\alpha}(E_p,-W) dW = \sum_{j',j} p_{j'} P^{\alpha}_{j'j}(E_p + e_j) \neq 1$. This reflects the fact that the map in Eq.~\eqref{scatteringmap} is non-unital, or equivalently that its dual is not trace preserving \cite{Rastegin2013,Rastegin2014,Goold2015,Goold2021}; unitality would require $\sum_{j} P^{\alpha}_{j'j}(E_p + e_j) = 1$ which is generally not obeyed in quantum scattering theory. Below, we show that both non-unitality and unitality are general features of the scattering process and discuss the physical conditions where each arises.

From our main result in Eq.~\eqref{fluctuationrelation} we can obtain an integral fluctuation relation $\int e^{-\beta W} P^{\alpha}(E_p,W) dW = \gamma^{\alpha}(E_p)$. Using the fact that $[\hat{H}_S,\Phi^{\alpha}(E_p)(\hat{\mathbb{I}}_S)]=0$ (which follows from the properties of the eigenoperators as shown in Appendix \ref{app:eigenoperators}) we can recast the normalization of the dual distribution as $\gamma^{\alpha}(E_p) = \int \mathbb{P}^{\alpha}(E_p,-W) dW = Z^{-1} \mathrm{Tr}_S\big[e^{-\beta \hat{H}_S}\Phi^{\alpha}(E_p)(\hat{\mathbb{I}}_S)\big] = Z^{\alpha}(E_p)/Z$, where $Z^{\alpha}(E_p) = \mathrm{Tr}[e^{-\beta \hat{H}_S^{\alpha}(E_p)}]$ is the partition function associated with a new system Hamiltonian $\hat{H}_S^{\alpha}(E_p) \equiv \hat{H}_S - \beta^{-1} \log \Phi^{\alpha}(E_p)(\hat{\mathbb{I}}_S)$ which depends on the dynamical map itself. The integral fluctuation relation then reads $\int e^{-\beta W} P^{\alpha}(E_p,W) dW = e^{-\beta \Delta F^{\alpha}(E_p)}$,
where $\Delta F^{\alpha}(E_p) = -\beta^{-1} \log [Z^{\alpha}(E_p)/Z]$ with $Z^{\alpha}(E_p)/Z = \gamma^{\alpha}(E_p)$ describes the free energy available from the non-unitality of the process; by its definition, it evidently vanishes for unital maps. Through Jensen's inequality, we obtain the following lower bound for the average energy change
\begin{align}
    \label{inequality}
    \langle W \rangle^{\alpha} (E_p) \geq \Delta F^{\alpha}(E_p) \; ,
\end{align}
where $\langle W \rangle^{\alpha} (E_p) \equiv \int W P^{\alpha}(E_p,W) dW$. The sign of the lower bound is determined by the sign of the quantity $\eta^{\alpha}(E_p) \equiv \gamma^{\alpha}(E_p) - 1$. When $\eta^{\alpha}(E_p) > 0$ the lower bound in Eq.~\eqref{inequality} becomes negative and an initially thermal system can release energy in the collision; while when $\eta^{\alpha}(E_p) \leq 0$ it is impossible to extract energy from the system, with the equality holding for unital dynamics. We show in Appendix~\ref{app:derivationeta} that $\eta^{\alpha}(E_p)$ can be written in the exact form
\begin{align}
    \label{eta}
    \eta^{\alpha}(E_p) & = \sum_{\Delta > 0} \tanh{\bigg(\frac{\beta \Delta}{2} \bigg)}\sum_{\substack{j',j:\\ e_{j'}-e_j = \Delta}} \Bigg( \frac{Z_{j'j}}{Z}\Bigg) \nonumber \\
    & \times [P^{\alpha}_{jj'}(E_p + e_{j'}) - P^{\alpha}_{j'j}(E_p + e_{j})] \; .
\end{align}
The first sum in the last expression is over all the energy gaps of the system. For a given energy gap $\Delta > 0$, the second sum is over all pairs of energy levels whose difference is $\Delta$ and $Z_{j'j} = e^{-\beta e_{j'}} + e^{-\beta e_{j}}$ is the partition function of one of these pairs. The last term describes the imbalance between relaxation and excitation probabilities of the pair, being positive (negative) when the former are higher (lower) than the latter. 

\textit{Discussion and example. ---} In general, it is difficult to predict the behaviour of $\langle W \rangle^{\alpha}(E_p)$ and $\eta^{\alpha}(E_p)$, since they depends strongly on the scattering matrix, which in turns depends on the system Hamiltonian $\hat{H}_S$ and scattering potential $\hat{V}(\hat{x})$. Whenever there is access to the multichannel scattering matrix (or collision cross section), such as in ultracold atom experiments (see e.g. Refs.~\cite{Schmidt2019,Widera2020}), these quantities can be determined. However, we can study their behaviour more generally based on universal scattering features in two regimes: when the kinetic energy is comparable to the minimum energy gap of the system or when it is much larger than the maximum energy gap. As an example, consider a particle colliding with a two-level system with energy gap $\Delta > 0$. For simplicity, we consider a spatially symmetric potential $\hat{V}(\hat{x}) = V(-\hat{x})$ in which case the scattering process is independent of the initial direction of the particle $\alpha$ \cite{Taylor2006,Jacob2021} and we omit this label. The relevant quantities in Eq.~\eqref{inequality} then read exactly
$\langle W \rangle (E_p) = (\Delta/2) \cosh^{-1}(\beta \Delta/2) [P_{10}(E_p + e_0)e^{\beta \Delta/2} - P_{01}(E_p + e_1) e^{-\beta \Delta/2}]$ and $\eta(E_p) = \tanh(\beta \Delta/2) [P_{01}(E_p + e_1) - P_{10}(E_p + e_0)]$, where $\ket{1}$ and $\ket{0}$ are the excited and ground state. 

\begin{figure}[t!]
\centering
\begin{subfigure}
  \centering
    \includegraphics[width=0.45\textwidth]{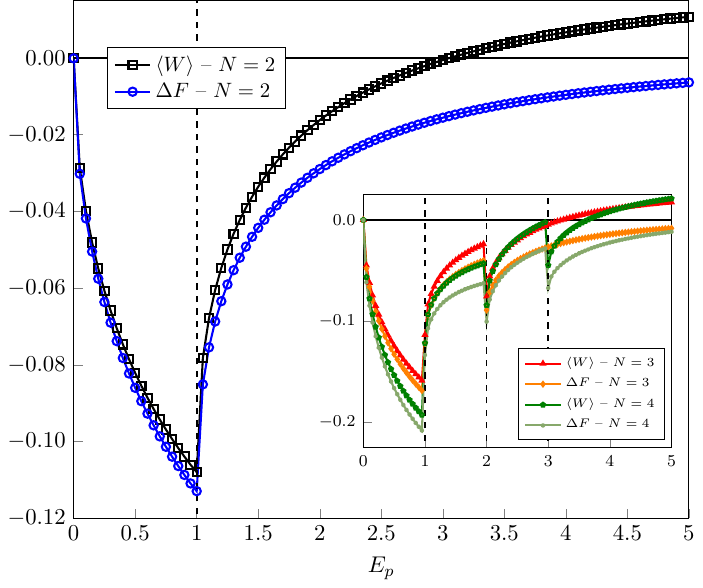}
\end{subfigure}
\begin{subfigure}
  \centering
  \includegraphics[width=0.47\textwidth]{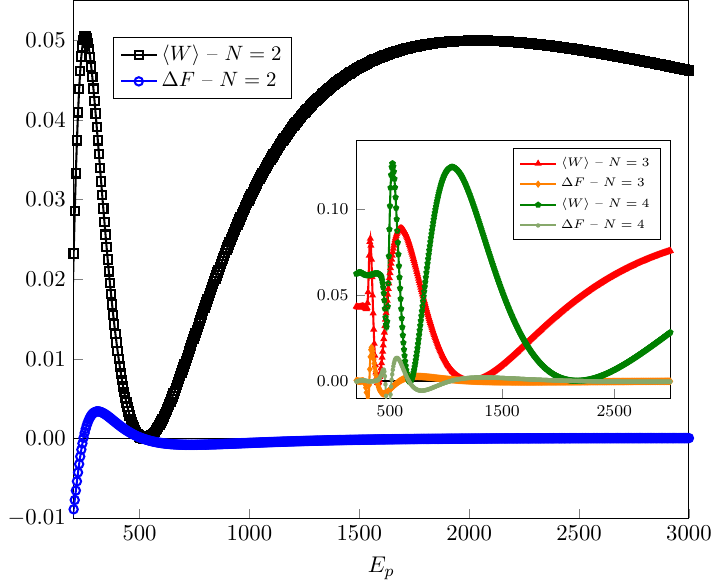}
\end{subfigure}
\caption{\label{fig-plots} Average system energy change and lower bound in Eq.~\eqref{inequality} at low and high kinetic energies (upper and lower panel, respectively). The explicit dependence of these quantities on $E_p$ was removed in the labels for simplicity. We consider a two-level ($N=2$) system $\hat{H}_S = (\Delta/2) \hat{\sigma}_{z}$ and scattering potential $\hat{V}(\hat{x}) = (V_0 \pi/2) \hat{\sigma}_x \otimes \cos(\pi \hat{x}/a)$, where $\Delta$ is the energy gap, $\hat{\sigma}_{z,x}$ are Pauli matrices and $V_0, a$ are the energy and length of the potential. The insets show the results when we add one and two more levels ($N=3$ and $N=4$), where the energies of the new levels are chosen such that the set of gaps for an $N$-level system is $\{\Delta, 2 \Delta,... (N-1)\Delta \}$; the interaction in the insets is $\hat{V}(\hat{x}) = (V_0 \pi/2) \hat{V} \otimes \cos(\pi \hat{x}/a)$ with $\hat{V}$ having 0s in the diagonal and 1s everywhere else. The scattering matrix is found by solving numerically the multichannel scattering equations \cite{Razavy2003,Jacob2021}. The parameters shown are $\Delta = m = a = 1$, $\beta = 0.1$, $V_0 = 100$ and the vertical dashed line in the upper panel indicates $E_p = \Delta, 2 \Delta, ... (N-1)\Delta$.}
\end{figure}

At low kinetic energies $0 \leq E_p < \Delta$ we see that Eq.~\eqref{inequality} allows for energy extraction from the system (Fig.~\ref{fig-plots}, upper panel). This is because system excitation is forbidden when the particle has an initial kinetic energy lower than the gap: the excitation channel is closed, i.e. $P_{10}(E_p + e_0) = 0$ for $0 \leq E_p < \Delta$. In contrast, a system initially at finite temperature has a non-zero probability to be excited and then relax in the collision: the relaxation channel is always open i.e. $P_{01}(E_p + e_1) \geq 0$ for $E_p \geq 0$. In this regime, we can then write $\langle W \rangle (E_p) = -\Delta f(\beta \Delta)P_{01}(E_p + e_1) \leq 0$ where $f(x) = (1 + e^{x})^{-1}$ is the Fermi function and $\eta(E_p) = \tanh(\beta \Delta/2) P_{01}(E_p + e_1) \geq 0$. Thus, the maximum energy that can be extracted from a two-level system in any scattering process is $\langle W \rangle_{\mathrm{ext}}^{\mathrm{max}} = \Delta f(\beta \Delta)$. Note when both channels are open, energy can still be extracted in the range $\Delta \leq E_p \leq E^{\mathrm{max}}_{p}$ where $E^{\mathrm{max}}_{p}$ is implicitly defined by $\langle W \rangle (E^{\mathrm{max}}_{p}) = 0$. Similar conclusions can be drawn for an $N$-level system at low kinetic energies, with different expressions for $\langle W \rangle (E_p)$ and $\eta(E_p)$, provided that we consider the minimum energy gap of the system. These were confirmed numerically (see inset in upper panel of Fig.~\ref{fig-plots}), where it is evident that a larger number of levels allows for more energy extraction at low energies. A systematic study of energy extraction for larger (many-body) systems is left for future work.

At high kinetic energies $E_p \gg \Delta$ energy extraction becomes impossible and we recover unital dynamics (Fig.~\ref{fig-plots}, lower panel). Since in this regime $\eta(E_p) \rightarrow 0$, the signature of unital dynamics is $P_{01}(E_p + e_1) = P_{10}(E_p + e_0)$ and we can write $\langle W \rangle (E_p) = \Delta \tanh(\beta \Delta/2) P_{10}(E_p + e_{0}) \geq 0$. The maximum energy consumed by the two-level system in any scattering process can never exceed $\langle W \rangle_{\mathrm{cons}}^{\mathrm{max}} = \Delta \tanh(\beta \Delta / 2)$. In fact, the convergence towards unitality at high kinetic energies is a universal feature of the scattering process, where the behaviour of the scattering matrix is mainly determined by the kinetic energy and depends weakly on the system energy gaps $P_{10}(E_p + e_{0}) = P_{10}(E_p + e_{1} - \Delta) \simeq P_{10}(E_p + e_{1}) = P_{01}(E_p+e_{1})$, where the last equality follows from the time-reversal symmetry of the scattering matrix \cite{Taylor2006,Jacob2021,Jacob2023} (see also Appendix~\ref{app:propertiesscatt}). Similar conclusions hold for an $N$-level system, with a different expression for $\langle W \rangle (E_p)$, provided that kinetic energy is much larger than the maximum energy gap. These conclusions also hold for non-symmetric potentials, since at sufficiently high kinetic energies the precise shape of the potential $\hat{V}(\hat{x})$ is not captured by the scattering matrix \cite{Jacob2022}. We confirmed numerically these predictions for larger system sizes (see inset in lower panel of Fig.~\ref{fig-plots}) and non-symmetric potentials (not shown).

Note that for a two-level system, the maximum energy that can be extracted $\langle W \rangle_{\mathrm{ext}}^{\mathrm{max}} = \Delta f(\beta \Delta)$ is maximal $\Delta/2$ at $\beta = 0$ (infinite temperature) and decreases monotonically to zero as $\beta \rightarrow \infty$ (zero temperature), while the maximum energy consumed $\langle W \rangle_{\mathrm{cons}}^{\mathrm{max}} = \Delta \tanh(\beta \Delta / 2)$ is zero at $\beta = 0$ and increases monotonically to $\Delta$ at $\beta \rightarrow \infty$. Curiously, there is a temperature above which extraction supersedes consumption $0 \leq \beta \leq \beta_0$, where $\beta_{0} = \Delta^{-1}\log(2)$ is determined by the intersection of both functions. At this threshold temperature we have $\langle W \rangle_{\mathrm{ext}}^{\mathrm{max}} = \langle W \rangle_{\mathrm{cons}}^{\mathrm{max}} = \Delta / 3$.

\textit{Conclusions. ---} We have shown how energy fluctuations of a quantum system can be studied within scattering theory beyond the macroscopic source limit. When a collision with a particle pushes the system away from thermal equilibrium, the probability distribution for the energy changes obeys a universal fluctuation relation \eqref{fluctuationrelation} which allows for energy releasing processes as dictated by non-unital dynamics. Such processes are particularly important if the kinetic energy of the particle is of the order of the energy fluctuations, highlighting the importance of non-unital maps in describing interactions with microscopic sources. At high kinetic energies, unitality is recovered, together with the standard fluctuation theorems for unital dynamics. 

Our results may surprise readers familiar with the second law of thermodynamics. As stated by Thomson and Planck: "There is no physical process whose sole effect is energy extraction from a thermal bath." However, we have to note that the state of the particle -- generally described by a wavepacket -- will be distorted in the scattering process \cite{Jacob2021,Jacob2023}. In this regard, we show in Appendix~\ref{app:EP} that the entropy production, defined as the average log-ratio of the probability for the forward process and the backward process, is always positive -- even at low kinetic energies, where energy extraction from a thermal system is possible. 
In Appendices \ref{app:unconditionedmap} and \ref{app:heatfluctuations} we prove that heat fluctuation theorems also follow from \eqref{fluctuationrelation} when the kinetic energy of the particle is thermally distributed. Our work provides a unifying perspective on thermodynamics of quantum systems within a realistic scattering setup.


\appendix

\section{Fluctuations for quantum maps \label{app:nonunital}}

We review here, in full generality, how to define energy fluctuations for dynamical maps, highlighting the importance of non-unital and unital maps. 

A quantum (or dynamical) map $\Lambda$ is a completely positive and trace preserving map, taking quantum states to quantum states $\hat{\rho}' = \Lambda(\hat{\rho})$ \cite{Kraus1971,Kraus1983,Nielsen2000,Alicki2007}. The dual (or adjoint) map $\Lambda^{\dagger}$ associated to $\Lambda$ is defined through $\mathrm{Tr}[\hat{O} \Lambda(\hat{\rho})] = \mathrm{Tr}[ \Lambda^{\dagger}(\hat{O}) \hat{\rho}]$, where $\hat{O}$ is an arbitrary linear operator. When $\hat{O} = \hat{\mathbb{I}}$ we get $\mathrm{Tr} \big[ \Lambda(\hat{\rho}) \big] = \mathrm{Tr} \big[\Lambda^{\dagger}(\hat{\mathbb{I}})\hat{\rho} \big] = \mathrm{Tr}[ \hat{\rho}]$ which follows from the fact that $\Lambda$ is trace preserving. Thus, we conclude that $\Lambda$ is trace preserving if and only if its dual is a unital map $\Lambda^{\dagger}(\hat{\mathbb{I}}) = \hat{\mathbb{I}}$, i.e. if its dual preserves the identity. However, note that $\Lambda$ itself is generally non-unital which means that its dual is not trace preserving: $\mathrm{Tr}[ \Lambda^{\dagger}(\hat{\rho})] = \mathrm{Tr}[ \Lambda(\hat{\mathbb{I}})\hat{\rho}] \neq \mathrm{Tr}[\hat{\rho}]$.

A dynamical map represents the most general type of evolution for an open quantum system \cite{Breuer2007,Rivas2012}. The map $\Lambda$ and its dual $\Lambda^{\dagger}$ can always be written as
\begin{align}
    \Lambda(\cdot) = \sum_{l} \hat{K}_l \cdot \hat{K}_l^{\dagger} \label{dynamicalmap} \\
    \Lambda^{\dagger}(\cdot) = \sum_{l} \hat{K}_l^{\dagger} \cdot \hat{K}_l
\end{align}
where $\{\hat{K}_l\}$ are called Kraus operators and the trace preserving property reads $\Lambda^{\dagger}(\hat{\mathbb{I}}) = \sum_l \hat{K}^{\dagger}_l \hat{K}_l = \hat{\mathbb{I}}$. In order to define energy fluctuations for a general dynamical map, we consider two (non-degenerate) Hamiltonians $\hat{H} = \sum_{n} E_n \hat{\Pi}_n$ and $\hat{H}' = \sum_{m} E'_m \hat{\Pi}'_m$ describing the energy of the quantum system before and after the open system evolution induced by Eq.~\eqref{dynamicalmap}, with $\hat{\Pi}_n =\ket{E_n} \bra{E_n}$ and $\hat{\Pi}'_m = \ket{E'_m} \bra{E'_m}$ being projectors onto the energy eigenbasis. According to the two-point measurement scheme \cite{Talkner2007a,Esposito2009,Campisi2011}, we measure the energy of the system before and after the evolution. We can then define the following probability distribution for the energy changes \cite{Rastegin2014,Goold2015,Goold2021}
\begin{align}
    \label{probabilitydistributionmaps}
    P(W) & = \sum_{l,m,n} \delta(W- E'_m + E_n) \mathrm{Tr}[\hat{\Pi}'_m \hat{K}_l \hat{\Pi}_n \hat{\rho} \hat{\Pi}_n \hat{K}_l^{\dagger} \hat{\Pi}'_m] \nonumber \\
    & = \sum_{l,m,n} \delta(W- E'_m + E_n) |\braket{E'_m|K_l|E_n}|^2 p_n \; ,
\end{align}
where $\rho$ is the initial quantum state of the system and $p_n \equiv \braket{E_n|\rho|E_n}$. Since the dynamical map is trace preserving, this distribution can be easily shown to be normalized $\int P(W) dW = \mathrm{Tr} [\sum_{l} \hat{K}^{\dagger}_l \hat{K}_l \hat{\rho}] = 1$. Consider now that the system is in a thermal state with respect to the initial Hamiltonian $\rho = e^{-\beta H} / Z$ where $Z = \mathrm{Tr}[e^{-\beta H}]$ is the partition function of the initial Hamiltonian. Then $p_n = (p_n/p'_m) p'_m = e^{\beta(E'_m - E_n)} e^{-\beta \Delta F} p'_m$, where $p'_m \equiv \braket{E'_m|\rho'|E'_m}$ and $\rho' = e^{-\beta H'} / Z'$ is the thermal state with respect to the final Hamiltonian, $Z' = \mathrm{Tr}[e^{-\beta H'}]$ is the associated partition function and $\Delta F = -\beta^{-1} \log(Z'/Z)$ is the free energy difference between the two thermal states. Using this in Eq.~\eqref{probabilitydistributionmaps} leads immediately to the relation
\begin{align}
    \label{fluctuationrelationmaps}
    e^{-\beta W} P(W) = e^{-\beta \Delta F} \mathbb{P}(-W) \; ,
\end{align}
where the dual distribution is defined by
\begin{align}
    \mathbb{P}(-W) & = \sum_{l,m,n} \delta(-W - E_n + E'_m) \mathrm{Tr}[\hat{\Pi}_n \hat{K}_l^{\dagger} \hat{\Pi}'_m \hat{\rho}' \hat{\Pi}'_m \hat{K}_l \hat{\Pi}_n] \nonumber \\
    & = \sum_{l,m,n} \delta(-W- E_n + E'_m) p'_m |\braket{E'_m|K_l|E_n}|^2 \; .
\end{align}
Note that this distribution encodes the energy changes in the reverse process, where the evolution is now induced by the dual map. However, the dual distribution is not normalized $\int \mathbb{P}(-W) dW = \mathrm{Tr} [\sum_{l} \hat{K}_l \hat{K}^{\dagger}_l \hat{\rho}'] = \mathrm{Tr}[\Lambda^{\dagger}( \hat{\rho}')] \equiv \gamma \neq 1$ since the dynamical map is not unital. Integrating Eq.~\eqref{fluctuationrelationmaps} leads to the modified Jarzynski equality \cite{Goold2021}
\begin{align}
    \int e^{-\beta W} P(W) dW = e^{-\beta [\Delta F - \beta^{-1}\log \gamma]} \; .
\end{align}
An application of Jensen's inequality for exponential functions $\langle e^{-X} \rangle \geq e^{-\langle X \rangle}$, where $\langle \cdot \rangle$ denotes the average with respect to some probability distribution, leads to the bound
\begin{align}
    \langle W \rangle \geq \Delta F -\beta^{-1} \log \gamma \; .
\end{align}
Note that in a cyclic process, the initial and final Hamiltonians are the same and thus $\Delta F = 0$. The scattering process studied here constitutes a cyclic process.


\section{Explicit form of Eq. 1 \label{app:derivationmap}}

The explicit evaluation of Eq.~\eqref{map} is best performed by changing from momentum eigenstates to kinetic energy eigenstates. We can do this through the resolution of identity in the Hilbert space of the particle
\begin{align}
    \hat{\mathbb{I}}_P = \int^{+\infty}_{-\infty} dp \ket{p}\bra{p} = \int^{+\infty}_{0} dE_p \sum_{\alpha = \pm} \ket{E^{\alpha}_p}\bra{E^{\alpha}_p} \; ,
\end{align}
where we separated the integral in momentum into its positive ($\alpha = +$) and negative ($\alpha = -$) contributions and then changed variables from momentum to kinetic energy. The kinetic energy eigenstates are then defined by $\ket{E^{\alpha}_{p}} \equiv \sqrt{|dp/dE_p|} \ket{\alpha|p|}$ where $|p| = \sqrt{2mE_p}$ and $|dp/dE_p| = \sqrt{m/|p|}$. They are orthogonal and obey $\braket{E^{\alpha}_p | E^{\beta}_q} = \delta_{\alpha\beta} \delta(E_p - E_q)$. Using these eigenstates, the map in Eq.~\eqref{map} can be written explicitly
\begin{align}
    \label{mapappendix}
    \Phi(\hat{\rho}_S) & = \int dE_p \int dE_q \int dE_{q'} \sum_{\beta,\beta'} \braket{E^{\beta}_q | \hat{\rho}_P | E^{\beta'}_{q'}} \nonumber \\
    & \sum_{\alpha}\braket{E^{\alpha}_p|\hat{S}|E^{\beta}_{q}} \hat{\rho}_S \braket{E^{\alpha}_p|\hat{S}|E^{\beta'}_{q'}}^{\dagger} \; ,
\end{align}
where we omit the integration limits and summation values for simplicity. The scattering operator in this representation can now be written as
\begin{align}
    \braket{E^{\alpha}_p|\hat{S}|E^{\beta}_{q}} & = \sum_{j',j} \ket{j'}\bra{j} \braket{E^{\alpha}_{p},j'|\hat{S}|E^{\beta}_{q},j} \nonumber \\
    & = \sum_{j',j} \ket{j'}\bra{j} \delta(E_p + e_{j'} - E_q - e_j) s^{\alpha \beta}_{j'j}(E_q + e_j) \nonumber \\
    & = \sum_{\Delta} \delta(E_p - E_q + \Delta) \sum_{\substack{j',j:\\ e_{j'}-e_j = \Delta}} s_{j'j}^{\alpha \beta}(E_q + e_j) \ket{j'}\bra{j} \nonumber \\
    & = \sum_{\Delta} \delta(E_p - E_q + \Delta) \hat{S}^{\alpha \beta}_{\Delta}(E_q) \; .
\end{align}
In the first line, we inserted two resolutions of identity in the Hilbert space of the system $\hat{\mathbb{I}}_S = \sum_{j} \ket{j}\bra{j}$; in the second line, we used the representation of the scattering operator in the eigenbasis of $\hat{H}_0$ \cite{Belkic2004,Taylor2006}; in the third line, we split the sum over system eigenstates into energy differences; in the last line, we defined the eigenoperators of the system which appear in Eq.~\eqref{eigenoperators}. Plugging this representation into Eq.~\eqref{mapappendix} and performing the integrals over $E_p$ and $E_{q'}$ yields
\begin{align}
    \label{mapappendix2}
    \Phi(\hat{\rho}_S) & = \int dE_q \sum_{\beta,\beta'} \sum_{\Delta,\Delta'} \rho^{\beta \beta'}_P(E_q,E_q + \Delta'-\Delta) \nonumber \\
    & \sum_{\alpha}\hat{S}^{\alpha \beta}_{\Delta}(E_q) \hat{\rho}_S \hat{S}^{\alpha \beta'}_{\Delta'}(E_q + \Delta' - \Delta)^{\dagger} \; ,
\end{align}
where we used the notation $\rho^{\alpha \beta}_P(E_{p},E_q) \equiv \braket{E^{\alpha}_{p}|\hat{\rho}_P|E^{\beta}_{q}}$ for the state of the particle. If this state obeys condition \eqref{narrow}, i.e. $\rho^{\beta \beta'}_P(E_{q},E_q - \Delta + \Delta') \simeq \delta_{\beta \beta'} \delta_{\Delta\Delta'}~ \rho^{\beta}_P(E_{q})$, where $\rho^{\beta}_P(E_q) \equiv \rho^{\beta \beta}_P(E_{q},E_q)$ is the kinetic energy distribution for a particle traveling with direction $\beta$, then the map simplifies to
\begin{align}
    \label{mapappendix3}
    \Phi(\hat{\rho}_S) & = \int dE_q \sum_{\beta} \rho^{\beta}_P(E_q) \sum_{\alpha,\Delta}\hat{S}^{\alpha \beta}_{\Delta}(E_q) \hat{\rho}_S \hat{S}^{\alpha \beta}_{\Delta}(E_q)^{\dagger} \; ,
\end{align}
which is equivalent to Eq.~\eqref{scatteringmapfull}. Eqs.~\eqref{scatteringmap} and \eqref{scatteringmapoperation} are easily obtained by introducing a new integration over a continuous variable $W$ with the help of a $\delta$ function.

\section{Properties of the scattering matrix}
\label{app:propertiesscatt}

\subsection{Unitarity}

The unitarity of the scattering operator $\hat{S}^{\dagger}\hat{S} = \hat{S} \hat{S}^{\dagger} = \hat{\mathbb{I}}$, where $\hat{\mathbb{I}} = \hat{\mathbb{I}}_S \otimes \hat{\mathbb{I}}_P$, enforces two properties on the scattering matrix which we now briefly review, but can be consulted in more detail elsewhere \cite{Belkic2004,Taylor2006,Jacob2021}. The two properties can be obtained through the expressions 
\begin{align}
    \braket{E^{\alpha'}_{p'},j'|\hat{S}^{\dagger}\hat{S}|E^{\alpha}_{p},j} = \delta_{\alpha \alpha'} \delta_{j'j} \delta(E_p-E_{p'}) \\
    \braket{E^{\alpha'}_{p'},j'|\hat{S} \hat{S}^{\dagger}|E^{\alpha}_{p},j} = \delta_{\alpha \alpha'} \delta_{j'j} \delta(E_p-E_{p'}) \; .
\end{align}
By inserting resolutions of identity in the full Hilbert space between the two operators, using the representation of the scattering operator $\braket{E^{\alpha'}_{p'},j'|\hat{S}|E^{\alpha}_{p},j} = \delta(E_{p'} + e_{j'} - E_p - e_j) s^{\alpha' \alpha}_{j'j}(E_p + e_j)$ and integrating out one of the $\delta$ functions, we find that the scattering matrix obeys
\begin{align}
    \sum_{\beta,k} [s^{\beta \alpha'}_{kj'}(E)]^* s^{\beta \alpha}_{kj}(E) = \delta_{\alpha \alpha'} \delta_{j'j} \label{scattprop1}\\
     \sum_{\beta,k} s^{\alpha' \beta}_{j'k}(E) [s^{\alpha \beta}_{jk}(E)]^* = \delta_{\alpha \alpha'} \delta_{j'j} \label{scattprop2}\; ,
\end{align}
where $E = E_p + e_j$ is the total energy. These are the most important properties of the scattering matrix holding for any fixed total energy $E$. The transition probability $P^{\alpha' \alpha}_{j'j}(E) \equiv |s^{\alpha'\alpha}_{j'j}(E)|^2$ then obeys $\sum_{j',\alpha'} P^{\alpha' \alpha}_{j'j}(E) = \sum_{j,\alpha} P^{\alpha' \alpha}_{j'j}(E) = 1$. Thus $P^{\alpha' \alpha}_{j'j}(E)$ is a bistochastic matrix at any fixed total energy $E$.

\subsection{Time-reversal symmetry}

Another important property of the scattering matrix is time-reversal symmetry. Namely, if $H$ and $\hat{H}_0$ commute with the time-reversal operator $\Theta$, then scattering is invariant under time-reversal and the scattering matrix obeys
\begin{align}
    \label{timereversal}
    s^{\alpha' \alpha}_{j'j}(E) = s^{-\alpha -\alpha'}_{jj'}(E) \; ,
\end{align}
where we assume for simplicity of notation that the eigenstates of $\hat{H}_S$ are time-reversal invariant $\Theta\ket{j} = \ket{j}$ \cite{Belkic2004,Taylor2006,Jacob2021}. As a consequence of this symmetry the transition probabilities obey $P^{\alpha' \alpha}_{j'j}(E) = P^{-\alpha -\alpha'}_{jj'}(E)$.

\section{Properties of the Kraus operators \label{app:eigenoperators}}

\subsection{Eigenoperators}

As we have discussed in the main text, the Kraus operators in Eq.~\eqref{scatteringmapoperation} are eigenoperators of the system. Using their definition in Eq.~\eqref{eigenoperators}, it is easy to see that they obey
\begin{align}
    & \hat{S}^{\alpha' \alpha}_{\Delta}(E_p)^{\dagger} \hat{S}^{\alpha' \alpha}_{\Delta}(E_p) \nonumber \\
    & = \sum_{\substack{k,j,j':\\ e_{k}-e_{j'} = e_{k}-e_{j} =\Delta}} \ket{j'}\bra{j} [s^{\alpha' \alpha}_{kj'}(E_p + e_{j'})]^* s^{\alpha' \alpha}_{kj}(E_p + e_j) \label{eqD1}\\
    & \hat{S}^{\alpha' \alpha}_{\Delta}(E_p) \hat{S}^{\alpha' \alpha}_{\Delta}(E_p)^{\dagger} \nonumber \\
    & = \sum_{\substack{k',k,j:\\ e_{k'}-e_{j} = e_{k}-e_{j} =\Delta}} \ket{k'}\bra{k} s^{\alpha' \alpha}_{k'j}(E_p + e_j) [s^{\alpha' \alpha}_{kj}(E_p + e_{j})]^* \; .\label{eqD2}
\end{align}
However, assuming that $\hat{H}_S$ is non-degenerate, we have $e_{k} - e_{j'} = e_{k} - e_j \Leftrightarrow e_{j'} = e_{j} \Rightarrow j = j'$. Therefore \eqref{eqD1} and \eqref{eqD2} become
\begin{align}
    & \hat{S}^{\alpha' \alpha}_{\Delta}(E_p)^{\dagger} \hat{S}^{\alpha' \alpha}_{\Delta}(E_p) = \sum_{\substack{k,j:\\ e_{k}-e_{j} = \Delta}} P^{\alpha' \alpha}_{kj}(E_p + e_j) \ket{j}\bra{j} \\
    & \hat{S}^{\alpha' \alpha}_{\Delta}(E_p) \hat{S}^{\alpha' \alpha}_{\Delta}(E_p)^{\dagger} = \sum_{\substack{k,j:\\ e_{k}-e_{j} = \Delta}} \ket{k}\bra{k} P^{\alpha' \alpha}_{kj}(E_p + e_j) \; .
\end{align}
These last two expressions are diagonal in the eigenbasis of $\hat{H}_S$. When they are used to compute the probability distribution and its dual [Eqs.~\eqref{probabilitydistribution} and \eqref{dualprobabilitydistribution}], they imply that only on the populations of the system are relevant. Summing over all energy differences and final particle directions $\alpha'$ yields
\begin{align}
    & \sum_{\Delta, \alpha'} \hat{S}^{\alpha' \alpha}_{\Delta}(E_p)^{\dagger} \hat{S}^{\alpha' \alpha}_{\Delta}(E_p) = \Phi^{\alpha}(E_p)^{\dagger}(\mathbb{I}_S) = \mathbb{I}_S \; , \\   
    & \sum_{\Delta,\alpha'} \hat{S}^{\alpha' \alpha}_{\Delta}(E_p) \hat{S}^{\alpha' \alpha}_{\Delta}(E_p)^{\dagger} = \Phi^{\alpha}(E_p)(\mathbb{I}_S) \nonumber \\
    & = \sum_{k} \ket{k}\bra{k} \sum_{j,\alpha'}P^{\alpha' \alpha}_{kj}(E_p + e_j) \; .
\end{align}
The first line expression follows from the properties of the scattering matrix in Eqs.~\eqref{scattprop1} and \eqref{scattprop2} and expresses the trace preserving property of the dynamical map. The second line shows that $\Phi^{\alpha}(E_p)(\mathbb{I}_S)$ is diagonal in the energy eigenbasis and therefore commutes with the $\hat{H}_S$; however, it cannot be further simplified since in general $\sum_{j,\alpha'}P^{\alpha' \alpha}_{kj}(E_p + e_j) \neq 1$ [see discussion below Eqs.~\eqref{scattprop1} and \eqref{scattprop2}]. Therefore, the dynamical map is generally non-unital.

\subsection{Time-reversal symmetry}

In addition, the time-reversal symmetry of scattering matrix enforces a time-reversal symmetry of the eigenoperators. Using Eq.~\eqref{timereversal} in Eq.~\eqref{eigenoperators} immediately leads to
\begin{align}
    \label{eigenoperatortimereversal}
    \hat{S}^{\alpha' \alpha}_{\Delta}(E_p)^{\dagger} =  \hat{S}^{-\alpha -\alpha'}_{-\Delta}(E_p - \Delta) \; .
\end{align}
The physical interpretation is clear: the operator $\hat{S}^{\alpha' \alpha}_{\Delta}(E_p)$ induces a transition in the system with energy $\Delta$ -- the particle having initial kinetic energy $E_p$ and direction $\alpha$, and final kinetic energy $E_p -\Delta$ and direction $\alpha'$ -- while its adjoint $\hat{S}^{\alpha' \alpha}_{\Delta}(E_p)^{\dagger}$ induces the time-reversed transition -- the particle having initial kinetic energy $E_p - \Delta$ and direction $-\alpha'$, and final kinetic energy $E_p$ and direction $-\alpha$. Note that the initial kinetic of the time-reversed process $E_p -\Delta$ depends on the energy jump $\Delta$ induced in the forward process. Although our main results presented in the paper do not require \eqref{timereversal} or \eqref{eigenoperatortimereversal}, we show in Appendix \ref{app:mr} that they shed light on the connection between the dual map and time-reversibility when our scattering setup is microscopically reversible.

\section{Entropy production}
\label{app:EP}

The inequality in Eq.~\eqref{inequality} can be interpreted as an expression of the entropy production for the scattering process. Indeed, we can write
\begin{align}
    \beta \langle W \rangle^{\alpha}(E_p) & = \int \beta W P^{\alpha}(E_p,W) dW \nonumber \\
    & = \int \log \Bigg[ \frac{P^{\alpha}(E_p,W)}{\mathbb{P}^{\alpha}(E_p,-W)} \Bigg] P^{\alpha}(E_p,W) dW \nonumber \\
    & = \int \log \Bigg[ \frac{P^{\alpha}(E_p,W)}{\tilde{\mathbb{P}}^{\alpha}(E_p,-W)} \Bigg] P^{\alpha}(E_p,W) dW \nonumber \\
    & - \log \gamma^{\alpha}(E_p) \; ,
\end{align}
where in the last line we introduced the normalized dual distribution $\tilde{\mathbb{P}}^{\alpha}(E_p,-W) \equiv \mathbb{P}^{\alpha}(E_p,-W) / \gamma^{\alpha}(E_p)$ and used the normalization of $P^{\alpha}(E_p)$. Therefore we conclude
\begin{align}
    \label{entropyproduction}
    \Sigma^{\alpha}(E_p) & = \int \log \Bigg[ \frac{P^{\alpha}(E_p,W)}{\tilde{\mathbb{P}}^{\alpha}(E_p,-W)} \Bigg] P^{\alpha}(E_p,W) dW \nonumber \\
    & = \beta \langle W \rangle^{\alpha}(E_p) + \log \gamma^{\alpha}(E_p) \geq 0 \; .
\end{align}
The entropy production is a relative entropy \cite{Nielsen2000} between two distributions: the energy distribution induced by the map and the (normalized) energy distribution induced by the dual map.

\section{Derivation of Eq. 11}
\label{app:derivationeta}

In order to derive Eq.~\eqref{eta}, we start from its definition 
\begin{align}
    \eta^{\alpha}(E_p) & = \gamma^{\alpha}(E_p) - 1 = \int \mathbb{P}^{\alpha}(E_p,-W) dW - 1 \nonumber \\
    & = \sum_{j',j} \frac{e^{-\beta e_{j'}}}{Z} P^{\alpha}_{j'j}(E_p + e_j) - 1 \; .
\end{align}
Expressing unity as $1 = \sum_{j'j} P^{\alpha}_{j'j}(E_p + e_j) e^{-\beta e_{j}} Z^{-1}$ and splitting the sum into energy changes yields
\begin{align}
    \eta^{\alpha}(E_p) & = \frac{1}{Z}\sum_{\Delta}\sum_{\substack{j',j:\\ e_{j'}-e_{j} = \Delta}} P^{\alpha}_{j'j}(E_p + e_j)(e^{-\beta e_{j'}}-e^{-\beta e_{j}}) \nonumber \\
    & = -\frac{2}{Z} \sum_{\Delta}\sum_{\substack{j',j:\\ e_{j'}-e_{j} = \Delta}} e^{-\beta \Delta/2} \sinh{\Bigg( \frac{\beta \Delta}{2} \Bigg)} P^{\alpha}_{j'j}(E_p + e_j) e^{-\beta e_j} \nonumber \; .
\end{align}
In the last expression we expressed the difference in exponentials as a hyperbolic sine function. Now we split the sum over energy differences $\Delta$ into positive and negative contributions, corresponding to system excitation and relaxation, rewriting the expression as
\begin{align}
    \eta^{\alpha}(E_p) & = -\frac{2}{Z} \sum_{\Delta > 0} \sinh{\Bigg( \frac{\beta \Delta}{2} \Bigg)} \nonumber \\
    & \times \Bigg[ \sum_{\substack{j',j:\\ e_{j'}-e_{j} = \Delta}} e^{-\beta \Delta/2} P^{\alpha}_{j'j}(E_p + e_j) e^{-\beta e_j} \nonumber \\
    & - \sum_{\substack{j',j:\\ e_{j'}-e_{j} = -\Delta}} e^{\beta \Delta/2} P^{\alpha}_{j'j}(E_p + e_j) e^{-\beta e_j} \Bigg] \nonumber \; .
\end{align}
The first sum is now over energy gaps (positive by definition), the first term in parenthesis corresponding to excitation and the last one to relaxation. Permuting the labels in the last expression, it can be further is simplified to
\begin{align}
    \eta^{\alpha}(E_p) & = \frac{2}{Z} \sum_{\Delta > 0} \sinh{\Bigg( \frac{\beta \Delta}{2} \Bigg)} \sum_{\substack{j',j:\\ e_{j'}-e_{j} = \Delta}} \nonumber \\
    & [ P^{\alpha}_{jj'}(E_p + e_{j'}) e^{-\beta e_{j'}} e^{\beta \Delta/2}- P^{\alpha}_{j'j}(E_p + e_j) e^{-\beta e_{j}} e^{-\beta \Delta/2} ] \nonumber \; .
\end{align}
The last step involves noting that we can write the quantity $Z_{j'j} = e^{-\beta e_{j'}} + e^{-\beta e_{j}}$ in two equivalent ways, namely $Z_{j'j} = 2 \cosh(\beta \Delta /2) e^{-\beta e_{j}} e^{-\beta \Delta/2} = 2 \cosh(\beta \Delta /2) e^{-\beta e_{j'}} e^{\beta \Delta/2}$ which leads directly to Eq.~\eqref{eta}.

\section{Microscopic reversibility}
\label{app:mr}

The results presented in the main text are independent of whether or not our scattering setup is microscopically reversible; nevertheless, we show here the impact of microscopic reversibility on our results. In addition to the scattering matrix obeying time-reversal symmetry in Eq.~\eqref{timereversal}, microscopic reversibility holds when the potential is spatially symmetric or when the particle approaches the scattering region from the left or right with equal probability \cite{Jacob2021}. We will assume the latter, which amounts to considering that the distribution of the particle introduced in Eq.~\eqref{narrow} is given by $\rho^{\alpha}_P(E_p) = \rho_P(E_p)/2$. The map in Eq.~\eqref{scatteringmapfull} then reads simply $\Phi(\hat{\rho}_S) = \int dE_p~\rho_P(E_p)~ \Phi(E_p)(\hat{\rho}_S)$, where $\Phi(E_p)$ is now only conditioned on the kinetic energy of the particle. It can be expressed as an integral over quantum operations, just like in Eq.~\eqref{scatteringmap}, with the expression
\begin{align}
    \Phi(E_p,W)(\cdot) = \sum_{\Delta}\delta(W-\Delta)\frac{1}{2}\sum_{\alpha',\alpha}\hat{S}^{\alpha' \alpha}_{\Delta}(E_p)\cdot \hat{S}^{\alpha' \alpha}_{\Delta}(E_p)^{\dagger} \; 
\end{align}
and Eq.~\eqref{probabilitydistribution} now becomes
\begin{align}
    \label{probabilitydistributionmr}
    P(E_p,W) & = \mathrm{Tr}_S\big[\Phi(E_p,W)(\hat{\rho}_S)\big] \nonumber \\
    & = \sum_{j',j} \delta(W-e_{j'}+e_{j}) P_{j'j}(E_p + e_j) p_j \; ,    
\end{align}
where $P_{j'j}(E_p+e_j) = 1/2\sum_{\alpha,\alpha'} |s^{\alpha'\alpha}_{j'j}(E_p+e_j)|^2$ are the transition probabilities. On the other hand, using Eq.~\eqref{eigenoperatortimereversal} the dual quantum operation can now be written as
\begin{align}
    & \Phi(E_p,W)^{\dagger}(\cdot) = \sum_{\Delta}\delta(W-\Delta)\frac{1}{2}\sum_{\alpha',\alpha}\hat{S}^{\alpha' \alpha}_{\Delta}(E_p)^{\dagger}\cdot \hat{S}^{\alpha' \alpha}_{\Delta}(E_p) \nonumber \\
    & = \sum_{\Delta}\delta(W-\Delta)\frac{1}{2}\sum_{\alpha',\alpha}\hat{S}^{\alpha' \alpha}_{-\Delta}(E_p -\Delta)\cdot \hat{S}^{\alpha' \alpha}_{-\Delta}(E_p - \Delta)^{\dagger} \nonumber \\
    & = \Phi(E_p - W,-W)(\cdot) \; .
\end{align}
We see that if our setup is microscopically reversible, the dual operation $\Phi(E_p,W)^{\dagger}$ is given by $\Phi(E_p - W,-W)$, where the kinetic energy in the argument is now conditioned on the energy change $W$ [see discussion below Eq.~\eqref{eigenoperatortimereversal}]. The dual distribution in Eq.~\eqref{dualprobabilitydistribution} now becomes
\begin{align}
    \label{dualprobabilitydistributionmr}
    P(E_p-W,-W) & = \mathrm{Tr}_S\big[\Phi(E_p-W,-W)(\hat{\rho}_S)\big] \nonumber \\
    & = \sum_{j',j} \delta(-W-e_{j'}+e_{j}) P_{j'j}(E_p + e_{j} - W) p_{j} \; ,
\end{align}
and our main result in Eq.~\eqref{fluctuationrelation} takes the form
\begin{align}
    e^{-\beta W}P(E_p,W) = P(E_p - W,-W) \; .
\end{align}
In this microscopically reversible formulation, it is clear that the map becomes unital when $\Phi(E_p - W,-W) \simeq \Phi(E_p,-W)$. As we have already discussed, a sufficient condition for this to happen is that the scattering matrix depends very weakly on the energy gaps of the quantum system as is the case at very high kinetic energies.

\section{The unconditioned map}
\label{app:unconditionedmap}

The map in Eq.~\eqref{scatteringmapfull} can also be written as $\Phi(\hat{\rho}_S) = \int dW\Phi(W)(\hat{\rho}_S)$ where 
\begin{align}
    \Phi(W)(\hat{\rho}_S) = \int dE_p~\sum_{\alpha}\rho^{\alpha}_P(E_p)~ \Phi^{\alpha}(E_p,W)(\hat{\rho}_S)
\end{align}
is the quantum operation in Eq.~\eqref{scatteringmapoperation} integrated over the initial state of the particle. It gives rise to an unconditioned form of the distribution in Eq.~\eqref{probabilitydistribution} reading
\begin{align}
    \label{probabilitydistributionintegrated}
    P(W) & = \mathrm{Tr}_S\big[\Phi(W)(\hat{\rho}_S)\big] = \sum_{j',j} \delta(W-e_{j'}+e_{j}) \mathbb{S}_{j'j} p_j \; ,    
\end{align}
where 
\begin{align}
    \label{stochasticmatrix}
    \mathbb{S}_{j'j} = \int dE_p~\sum_{\alpha} P^{\alpha}_{j'j}(E_p + e_i)\rho^{\alpha}_P(E_p)
\end{align}
is a stochastic matrix ruling the transition probabilities for the system. For an initially thermal system, the fluctuation relation in Eq.~\eqref{fluctuationrelation} now reads 
\begin{align}
    \label{fluctuationrelationunconditioned}
    e^{-\beta W}P(W) = \mathbb{P}(-W)
\end{align}
where $\mathbb{P}(-W) = \mathrm{Tr}_S\big[\Phi(W)^{\dagger}(\hat{\rho}_S)\big] = \sum_{j',j} \delta(-W-e_{j}+e_{j'}) p_{j'}\mathbb{S}_{j'j}$ is the dual distribution integrated over the initial state of the particle. From this relation, we can derive an inequality analogous to \eqref{inequality} but unconditioned on the initial state of the particle. In summary, our results carry over for the unconditioned map in Eq.~\eqref{scatteringmapfull}.

\section{Heat fluctuation theorems}
\label{app:heatfluctuations}

The properties of the stochastic matrix $\mathbb{S}_{j'j}$ in Eq.~\eqref{stochasticmatrix} have been studied in Ref.~\cite{Jacob2021} in the context of thermalization. Namely, it was shown if the kinetic energy of the particle is thermally distributed $\rho^{\alpha}_P(E_p) \sim e^{-\tilde{\beta}E_p}$ with inverse temperature $\tilde{\beta}$, then detailed balance 
\begin{align}
    \mathbb{S}_{j'j} = e^{- \tilde{\beta} (e_{j'} - e_{j})} \mathbb{S}_{jj'}
\end{align}
holds, provided the scattering setup is microscopically reversible (see Appendix~\ref{app:mr}). In this case, applying detailed balance to Eq.~\eqref{probabilitydistributionintegrated} now yields 
\begin{align}
    \label{detailedbalance}
    P(W) = e^{- \tilde{\beta}W} \mathbb{P}(W) \; .
\end{align}
This new result expresses a detailed balance symmetry between the energy distribution and its dual, being independent of the state of the system. However, when the system is thermal then Eqs.~\eqref{fluctuationrelationunconditioned} and \eqref{detailedbalance} together imply
\begin{align}
    e^{-(\beta - \tilde{\beta}) W}P(W) = P(-W) \; ,
\end{align}
which are the heat exchange fluctuation theorems \cite{Jarzynski2004}.

\bibliographystyle{apsrev4-2}
\bibliography{biblio}

\end{document}